\documentclass[aps,pra,twocolumn,superscriptaddress,nofootinbib]{revtex4-2}

\usepackage{amsmath}
\usepackage{amssymb, bm}
\usepackage{amsfonts}
\usepackage{braket}
\usepackage{comment}
\usepackage{gensymb} 
\usepackage{graphicx}
\usepackage{mathtools, bm}
\usepackage{MnSymbol}
\usepackage{natbib}
\usepackage{xcolor}
\usepackage{xspace}
\usepackage{soul}

\newcommand{\rmi}{\mathrm{i}}
\newcommand{\rme}{\mathrm{e}}
\newcommand{\rmd}{\mathrm{d}}
\newcommand{\taucc}{\ensuremath{\tau_{\rm cc}}\xspace}
\newcommand{\tauxuvir}{\ensuremath{\tau}\xspace}
\renewcommand{\tauxuvir}{\ensuremath{\tau_{\mbox{\tiny ir}}}\xspace}
\newcommand{\ini}{\ensuremath{_0}\xspace}
\newcommand{\ie}{i.e.,\xspace}
\newcommand{\eg}{{\em eg}\xspace}

\newcommand{\XUV}{XUV\xspace}
\newcommand{\IR}{IR\xspace}
\newcommand{\RABBIT}{RABBIT\xspace}

\renewcommand{\Re}{\mathrm{Re}\,}
\renewcommand{\Im}{\mathrm{Im}\,}

\newcommand{\laser}{\ensuremath{_{\mbox{\tiny L}}}}
\renewcommand{\laser}{\ensuremath{_0}}

\newcommand{\wL}{\ensuremath{\omega\laser}\xspace}

\begin{document}

\title{Probing Wigner time delays with photoelectron interferometry: \\Anisotropic long-range imprint of the short-range centrifugal potential
}

\newcommand{\LCPMR}{Sorbonne Universit\'e, CNRS, Laboratoire de Chimie Physique-Mati\`ere et Rayonnement, LCPMR, F-75005 Paris, France}

\author{Morgan Berkane}
\author{Camille L\'ev\^eque}
\author{Richard Ta\"\i eb}
\email{richard.taieb@sorbonne-universite.fr}
\author{J\'er\'emie Caillat}
\author{Jonathan Dubois}
\affiliation{\LCPMR}
\date{\today}

\begin{abstract}
We  consider the two-photon ionization of Hydrogen-like atoms. 
We find an approximate expression of the long-range phase based on an asymptotic expansion of the continuum eigenfunctions within the Wentzel-Kramers-Brillouin approximation.
Combined with commonly used perturbative approaches, the resulting analytic formalism can treat, at the same time, the two-photon propensity rules, the anisotropy in the continuum-continuum photoionization time delay and the soft-photon regime.
\end{abstract}
\maketitle

\section{Introduction}

Attosecond spectroscopies rely to a large extent on interferometric schemes, due to the extremely short time scale and the wave-like nature of the investigated  quantum processes~\cite{Agostini2004, Krausz2009}. In particular, two pillars of attosecond science, \ie the `attosecond streaking'~\cite{Hentschel2001} and the `reconstruction of attosecond beatings by interferences of two-photon transitions' (\RABBIT)~\cite{Paul2001} techniques, have been used to revisit photoemission in the time domain~\cite{Cavalieri2007,Schultze2010,Klunder2011}. 

The attosecond streaking and RABBIT are schemes based on two-photon ionization with extreme ultra-violet (\XUV) attosecond pulses dressed with an infra-red (\IR) field. 
Using them as \XUV-pump \IR-probe approaches to gain temporal insight on photoemission from atoms~\cite{Schultze2010, Klunder2011}, through measurements of the photoelectron's spectral phase, triggered  intense experimental and theoretical activities, see \eg~\cite{Kheifets2023} and references therein. A major issue  addressed in this context concerns the intricate roles of the \XUV pump and the \IR probe and more specifically on the imprint of the latter on the measured `photoemission delays'. Analytical derivations based on universal asymptotic expansions of the wave functions~\cite{Schultze2010, Klunder2011,Dahlstrom2012,Dahlstrom2013} and classical arguments~\cite{Nagele2011, Nagele2012, Pazourek2015, Saalmann2020} have proved efficient for the modeling and interpretation in various studies over the past decade, see \eg Refs.~\cite{Worner2016,Bray2018, Berkane2024,Loriot2024}. 

However, these approximate derivations are intrinsically isotropic, in the sense that they cannot account for the angular variations of the correction between the measurements on the one hand and the probed dynamics on the other hand, in investigations of {\em orientation-resolved} photoemission~\cite{Hockett2017, Bray2018,Fuchs2020,Autuori2022,Berkane2024}. The probe-induced anisotropy is particularly striking in the seminal work published in Ref.~\cite{Heuser2015}, where \RABBIT measurements of atomic photoemission delays in He display significant angular variations, while the probed Wigner delay associated with single-photon ionization in this case is isotropic.

From a fundamental perspective, this asymmetry is related to the dipole selection rules in one- and two-photon processes and to the so-called Fano propensity rules~\cite{Fano1985} applied to two-photon ionization~\cite{Busto2019}. These  rules formalize an imbalance in the angular momentum distribution in the two `arms' of the \RABBIT interferometer, resulting in an angular redistribution of the modulus and phase of the photoemission probability amplitudes. By using standard asymptotic expansion of the continuum wave functions, the original analytical derivations of the probe influence in \RABBIT measurements of photoemission delays~\cite{Klunder2011, Dahlstrom2012,Pazourek2015} fail to account for these two-photon Fano propensity rules. 

In this article, we derive a semi-analytic alternative representation of the continuum wave functions which is sufficient to recover, at least qualitatively, the asymmetric influence of the probe in \RABBIT measurements. We use an expansion of the continuum wave functions up to the first order in $r^{-1}$, using the Wentzel-Kramers-Brillouin (WKB) approximation~\cite{Bethe1957}. In this framework, our semi-analytical expressions appear explicitly as an improvement over the standard derivations, and are sufficient to retrieve the Fano propensity rules. Moreover, the WKB approach, by its  semiclassical nature, is particularly suited to extract intuitive physical insight from the obtained analytical resuts. In Sec.~\ref{sec:formalism}, we review the formalism used in \RABBIT, in terms of phases, delays, transitions amplitudes and continuum wave functions. In Sec.~\ref{sec:anisotropy}, focused on the question of anisotropy, we present our approach and demonstrate its ability to reproduce the two-photon Fano propensity rules. In Sec.~\ref{sec:links}, we illustrate the consistency of our work with respect to a selection of formerly published studies. We furthermore exploit its potential to provide finer insight to state-of-the art modelling and interpretation of time-resolved photoemission measurements. We provide a summary and conclusions in Sec.~\ref{sec:conclusion}.

\section{Measuring photoemission delays with RABBIT \label{sec:formalism}}
\subsection{Measurable atomic delay}
\begin{figure}[t]
    \centering
    \includegraphics[width=.25\textwidth]{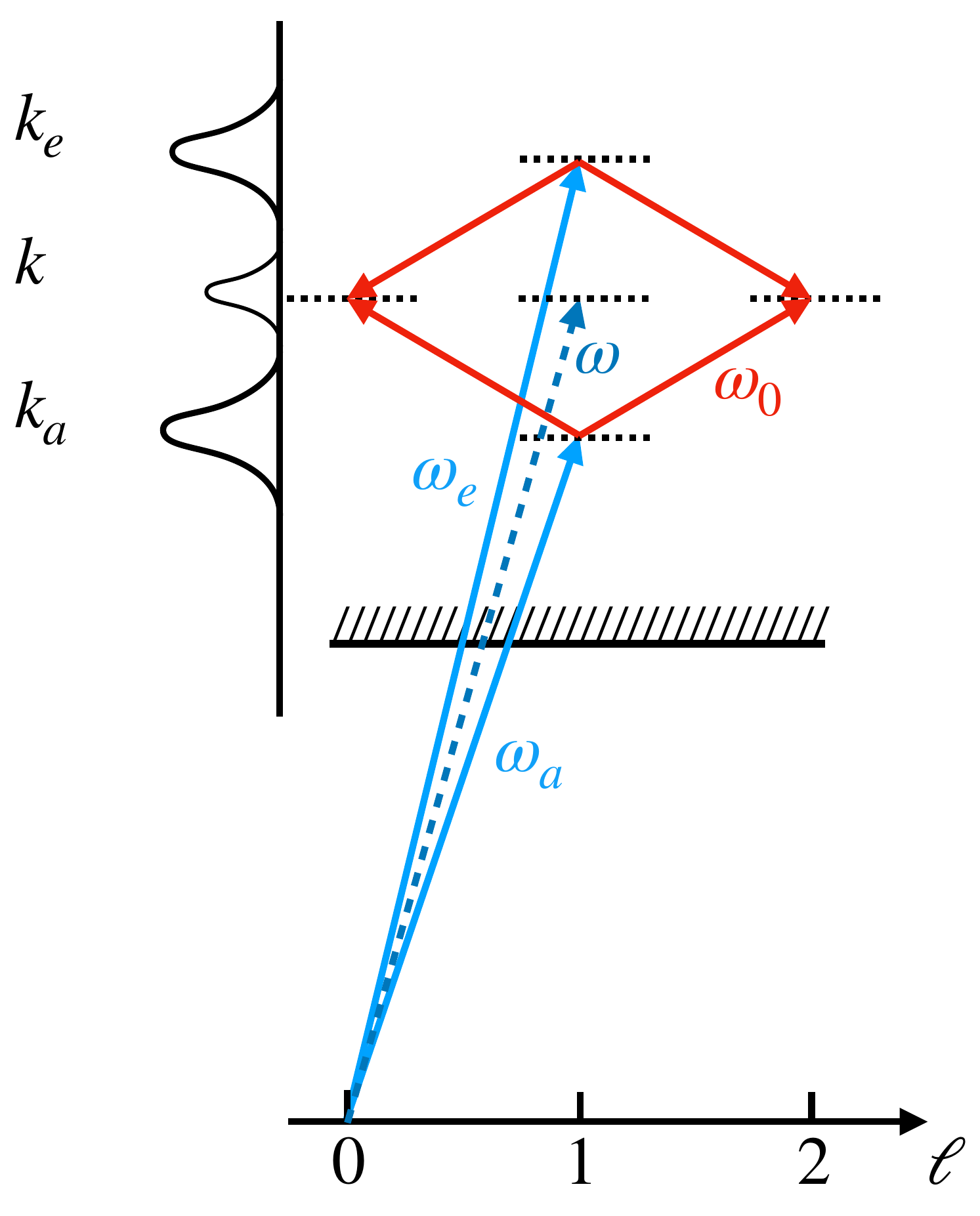}
    \caption{Principle of the RABBIT scheme with angular momentum resolution. The two-photon transitions are indicated by plain arrows. The dashed-line arrow show the virtual one-photon transition meant to be probed with the RABBIT scheme according to Refs.~\cite{Klunder2011, Dahlstrom2013}. See text for details.
    \label{fig:rabbitscheme}}
\end{figure}

In this paper, we focus on the temporal information encoded by a given sideband in a RABBIT scheme, as sketched in Fig.~\ref{fig:rabbitscheme} and summarized hereafter. 

We consider the photoemission of an atom by a pair of consecutive XUV odd harmonics (with frequencies $\omega_a$ and $\omega_e$) of an IR field, in presence of the fundamental field (typically 800 nm wavelength, which corresponds to a photon energy $\wL=1.55$ eV). The obtained photoelectron spectrum contains two main peaks separated by $2\wL$, corresponding to the absorption of each harmonic, and an additional sideband peak in between, resulting from two-photon transitions involving the harmonics and the IR field. The energy relationship between the considered harmonics and the fundamental IR field is such that two quantum paths lead to the sideband formation: absorption of the lowest harmonic and {\em absorption} of an IR photon (labelled `$a$' in Fig.~\ref{fig:rabbitscheme} and in the following), absorption of the highest harmonic and {\em emission} of an IR photon (`$e$').

When considering orientation-resolved measurements, the photoelectron momentum distribution (PMD) associated with the sideband is hence expressed as the coherent sum
\begin{eqnarray} \nonumber
    \mathcal{I}(\mathbf{k};\tauxuvir) &\propto& \left| M_a (\mathbf{k}) F_a \rme^{ - \rmi \varphi_a -\rmi \wL \tauxuvir} + M_e (\mathbf{k}) F_e \rme^{ - \rmi \varphi_e + \rmi \wL \tauxuvir} \right|^2. \\ \label{eq:Sperturbatif}
\end{eqnarray}
In this expression, $F_a$ and $F_e$ are the field strengths of the two harmonics, labelled according to the path to which they contribute, $\varphi_a$ and $\varphi_e$ are the associated phases, and $\wL\tauxuvir$ corresponds to the phase of the IR field, here written as a function of the experimentally tunable IR delay \tauxuvir with respect to the XUV. The factors $M_a (\mathbf{k})$ and $M_e (\mathbf{k})$ are the two-photon transition matrix elements associated with each path~\cite{Veniard1996,muller2002}, where $\mathbf{k}$ is the  photoelectron's asymptotic momentum vector. 

Investigating the dynamics of photoemission in RABBIT experiments consists in interpreting the so-called `atomic phase' corresponding to
\begin{eqnarray}\label{eqn:atomic_phase}
\Delta \phi_A(\mathbf{k})=\arg M_{a}(\mathbf{k}) - \arg M_{e}(\mathbf{k}).
\end{eqnarray}
This phase is accessible through a series of $\mathcal{I}(\mathbf{k};\tauxuvir)$ measurements scanning \tauxuvir and calibrated with respect to the harmonic phases~\cite{Klunder2011}. 
Various time-domain interpretations of the atomic phase have been proposed over the years, see \eg~\cite{Vacher2017} and references therein. Here we consider the approach first introduced in~\cite{Klunder2011} which consists in defining an `atomic delay'
\begin{subequations}
\begin{eqnarray}
\tau_A(\mathbf{k})&=&-\frac{\Delta\phi_A(\mathbf{k})}{2\wL}
\end{eqnarray}
and relating it to the Wigner delay~\cite{Wigner1955} 
\begin{eqnarray}\label{eq:tauW}
\tau_W(\mathbf{k})&=&\frac{1}{k}\frac{\partial \eta(\mathbf{k})}{\partial k}
\end{eqnarray}
associated with a virtual one-photon process leading to the sideband energy $E=k^2/2$, see dashed arrow in Fig.~\ref{fig:rabbitscheme}.
In expression~\eqref{eq:tauW}, $\eta(\mathbf{k})$ is the usual scattering phase shift of the photoelectron~\cite{Messiah1961} which result only from the system- and channel-specific short-range contributions. Any attempt to relate $\tau_A(\mathbf{k})$ to $\tau_W(\mathbf{k})$ implies that the continuum is smooth enough for $\eta(\mathbf{k})$ to vary almost linearly within the $2\wL$-wide spectral range separating the two consecutive harmonic peaks. Thus, $\tau_A(\mathbf{k})$ can be seen as a finite difference approximation of $\tau_W(\mathbf{k})$~\cite{Klunder2011}, up to a correction
\begin{eqnarray} \label{eq:tau_RABBIT}
    \taucc(\mathbf{k})  &=& \tau_{A} (\mathbf{k}) - \tau_W(\mathbf{k}).
\end{eqnarray}
\end{subequations}
In this context, RABBIT thus appears as an XUV-pump IR-probe scheme to access the dynamics of one-photon ionization in terms of Wigner delays. Yet, its full exploitation requires the knowledge of the correction term $\taucc(\mathbf{k})$.

Insights on this term is obtained by expanding the transition amplitudes $M_a(\mathbf{k})$ and $M_e(\mathbf{k})$ according to the second order perturbation theory, with an {\em ad hoc}  representation of the continuum wave functions. This is addressed in the following, with the objective to account for the anisotropy of $\taucc(\mathbf{k})$, \ie the anisotropy induced by the {\em probe} stage in the RABBIT scheme, a consequence of the two-photon transition selection rules {observed both in numerical simulations~\cite{Bray2018} and in experiments~\cite{Heuser2015}}. 

\subsection{Second-order perturbation theory}

 In order to focus on the anisotropy specifically brought by the two-photon processes rather than by the initial state or by the ionic potential (see~\cite{gaillac2016,Berkane2024} and references therein), we consider the case of a spherically symmetric $\ell=0$  initial state, and  linearly polarized colinear fields. By choosing the polarization direction as the quantization axis, the orbital magnetic quantum number $m=0$ is conserved during the processes. More importantly, a single angular-momentum channel ($\ell=1$) is open in one-photon ionization, \ie $\eta(\mathbf{k})=\eta_1(k)$ for all momentum directions $\hat{k}$, $k$ denoting the momentum magnitude. In contrast, two channels ($L=0,2$) are open in a two-photon process, see Fig.~\ref{fig:rabbitscheme}. 
 In other words, the dynamics of the {\em probed} process is isotropic, while the {\em probe} process -- and hence $\taucc(\mathbf{k})$ -- is anisotropic.

For each path ($\alpha=a,e$), the two-photon transition matrix element reads~\cite{Dahlstrom2013} 
\begin{equation}
\label{eq:Mtot}
    M_{\alpha} (\mathbf{k}) = \dfrac{(8 \pi)^{5/2}}{6 \, \rmi} \sum_{L = 0,2} C_{L 0} Y_{L 0} (\hat{k}) \rme^{\rmi \eta_{L} (k)} T^{\alpha}_{L} (k)
\end{equation}
where $C_{0 0} =1/2$ and $C_{20}={-}1/\sqrt{5}$ are the appropriate Clebsch-Gordan coefficients. $Y_{L 0}$ are the normalized spherical harmonics and $\eta_{L}$ are the scattering phases of the open channels $L=0,2$.  We emphasize that the phase of interest is ultimately $\eta_1(k)$, which does not appear explicitly in Eq.~\eqref{eq:Mtot} but emerges from the radial integrals $T^{\alpha}_{L} (k)$. The latter can be expressed as~\cite{muller2002}\footnote{All through the paper, the bra-ket notation designates the scalar product with respect to the radial coordinate $r$.}
\begin{equation}
\label{eq:Talphaell}
    T^{\alpha}_{L} (k)  = \bra{R_{kL}} r \ket{\rho_{k_{\alpha} 1}},
\end{equation}
where  $R_{kL}(r)$ is the radial component of the final state's (reduced) wave function in channel $L$ and $\rho_{k_{\alpha} 1}(r)$ the radial part of the so-called first-order perturbed wave function defined as
\begin{equation}
\label{eq:pertwf}
\rho_{k_{\alpha} 1}(r)
= \lim\limits_{\epsilon\rightarrow0^+}\sumint_{\nu} R_{\nu 1}(r) \dfrac{  \bra{R_{\nu 1}} r \ket{R\ini}}{E\ini + \omega_{\alpha} - E_\nu + \rmi \epsilon}.
\end{equation} 
All through the paper, we assume $E_0 {+} \omega_\alpha {>}0$ (with $E_0$ the initial energy and $\omega_\alpha$ is the XUV photon energy in the considered path) such that $\rho_{k_{\alpha} 1}$ behaves as a continuum wave function~\cite{Toma2002} with asymptotic momentum $k_\alpha=\sqrt{2(E_0{+}\omega_\alpha)}$. 
With these notations the time delay of the probed one-photon process [Eq.~\eqref{eq:tauW}] can be approximated as the finite difference 
\begin{equation}\label{eq:tauWapprox}
\tau_W (\mathbf{k}) = \frac{\eta_1 (k_e) - \eta_1 (k_a)}{2\wL}.
\end{equation}

In Eq.~\eqref{eq:pertwf},  $R\ini(r)$ is the radial part of the initial state, the sum/integral spans the $\ell=1$ manifold of the complete atomic spectrum, each of its state being associated with the radial wave function $R_{\nu 1}(r)$ and energy $E_\nu$. Note that, following the reasoning of Ref. \cite{Klunder2011}, we have discarded in Eq.~\ref{eq:pertwf} the term resulting from the paths where the infrared photon is absorbed first, which is negligible compared with the one considered here.

According to Eq.~\eqref{eq:pertwf},  the dominant contribution to $\rho_{k_{\alpha} 1}(r)$ in each path is proportional to the resonant $R_{k_\alpha 1}(r)$, \ie to the final wave function reached by the XUV alone in the probed process. This underlines the possibility to isolate $\eta_1(k_\alpha)$ in the phase of $M_\alpha(k)$.

\begin{figure*}
    \centering
    \includegraphics[width=.75\textwidth]{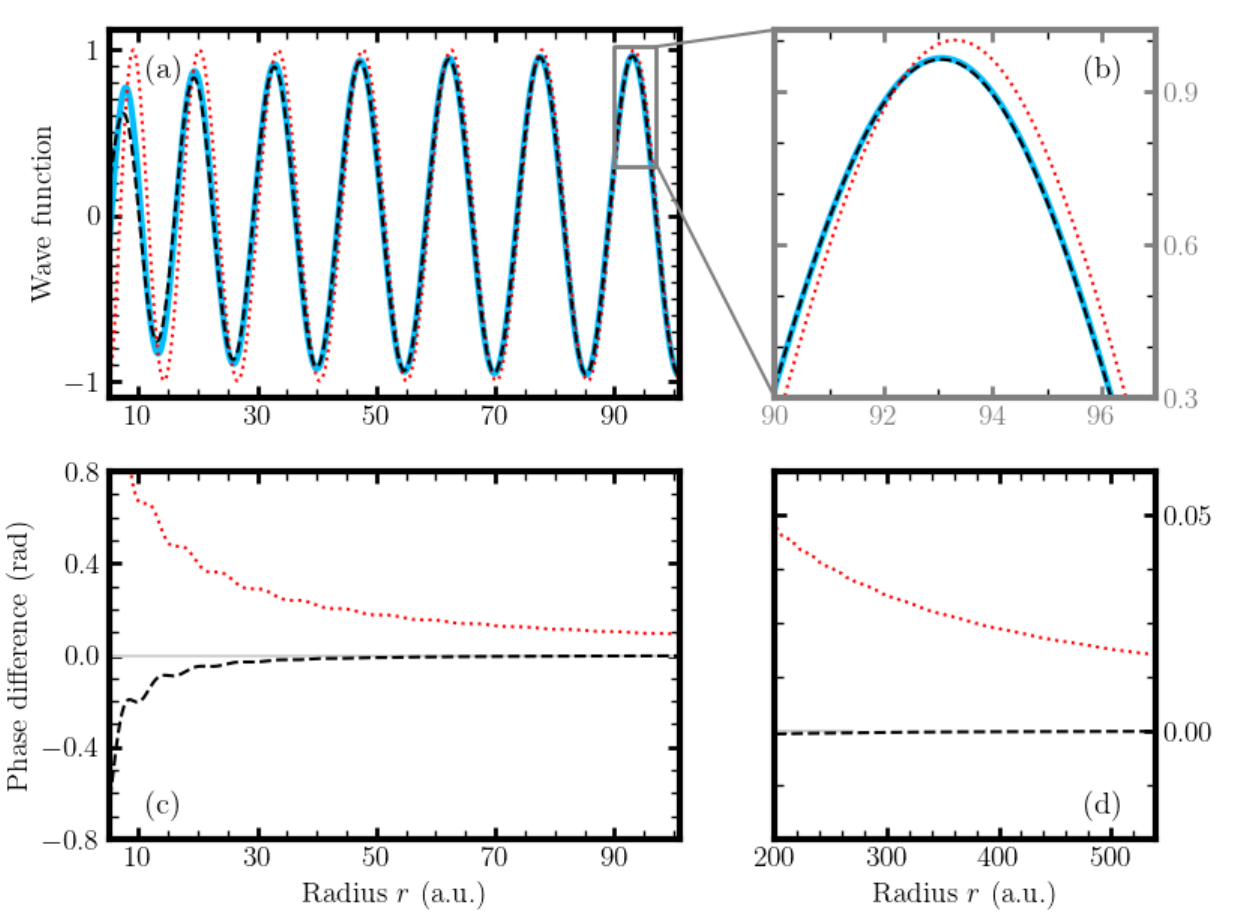}
    \caption{$L=0$ continuum state of the hydrogen atom at $k=0.37$ a.u. Upper pannels: Radial wave function obtained as the exact solution of the time-independent Schr\"{o}dinger equation (blue solid line), as the  approximation given by Eq.~\eqref{eq:WKB_psi} including $\beta_{\kappa \ell}$ (black dashed line), and as the standard approximation~\cite{Klunder2011, Dahlstrom2013} discarding $\beta_{\kappa \ell}$ (red dotted line).
    Lower panels: Phase difference between each of the two approximate wave functions (same color code as in the upper panels) and the exact one. The left and right columns highlight two different radius ranges.}
    \label{fig:WKB}
\end{figure*}

In Ref.~\cite{Klunder2011}, standard asymptotic expansions of $R_{kL}(r)$ and $\rho_{k_{\alpha}1}(r)$, \ie up to zeroth order in $r^{-1}$, are used. This approximation somehow formalizes a commonly accepted hypothesis, according to which the IR transition in the RABBIT scheme takes place {\em significantly} after the XUV absorption, \ie when the photoelectron is beyond the $L$-dependent short-range influence of ionic potential. Consequently, $T_{L}^{\alpha}$ is found independent of $L$ which eventually leads to an isotropic expression of $\tau_A$, and therefore of \taucc, as summarized in Appendix~\ref{sec:Klunder}. 

In the next section, we go beyond this approximation and present a semi-classical treatment of the two-photon transition amplitudes that provides a properly anisotropic~\cite{Heuser2015,Busto2019} semi-analytic derivation for \taucc.

\section{Accounting for the anisotropy}\label{sec:anisotropy}

In order to let the scattering phase difference $\eta_1(k_e)-\eta_1(k_a)$ emerge from the RABBIT phase, we explicitly express the total phase of the continuum radial wave functions $\rho_{k_\alpha,1}(r)$ and $R_{kL}(r)$ in the generic form
\begin{equation}\label{eqn:S}
        S_{\kappa \ell}(r) = \underbrace{\kappa r +  \dfrac{Z}{\kappa} \ln \left( 2 \kappa r \right)  - \dfrac{\ell \pi}{2} +  \eta_{\ell} (\kappa)}_{\equiv S^{\infty}_{\kappa \ell}(r)} + \Delta S_{\kappa\ell}(r).
\end{equation}
Here $\kappa$ is the asymptotic momentum norm and $\ell$ the angular momentum, $Z$ representing the charge of the photoelectron's parent ion.
The first terms encompassed in $S^{\infty}_{\kappa \ell}(r)$, including $\eta_\ell(\kappa)$, correspond to the universal {\em asymptotic} phase of an ionic radial continuum wave function. Only this expression was considered in the seminal works on RABBIT measurements of photoemission delays~\cite{Klunder2011,Dahlstrom2013} that led to an isotropic \taucc. Here, we explicitly added a term $\Delta S_{\kappa\ell}(r)$ to be determined, expected to include short-range effects and to account for the anisotropy of \taucc. This term vanishes at large $r$ such that
\begin{eqnarray}\label{eqn:Sasym}
S_{\kappa \ell}(r)\underset{r\to\infty} \sim S^{\infty}_{\kappa \ell}(r).
\end{eqnarray}

\subsection{WKB continuum wave functions}

To proceed, we chose the WKB formalism~\cite{Bethe1957}, which  provides both intuitive arguments in terms of classical mechanics and satisfactory quantitative results in related contexts, see \eg ~\cite{Zahn2023} and references therein. We furthermore restricted $\Delta S_{\kappa\ell}(r)$ to the first non-vanishing term in its expansion in powers of $r^{-1}$. Within WKB, the final continuum and intermediate first-order perturbed wave function are approximated as 
\begin{subequations}
\label{eq:WKB_formalism}
\begin{eqnarray}
    R_{kL} (r)
    &\simeq&
    \dfrac{\sin \big( S_{kL} (r) \big)}
    {\left[k^2 {-} 2 V_{L} (r)\right]^{1/4}}  ,\label{eq:WKB_psi} 
\end{eqnarray}
and
\begin{eqnarray}
    \rho_{k_\alpha 1}(r) &\simeq& \dfrac{\exp \big( \rmi S_{k_\alpha 1} (r) \big)}{\left[k_\alpha^2 {-} 2 V_{1} (r)\right]^{1/4}}  ,
\end{eqnarray}
respectively, where $V_\ell(r)$ is the channel-dependent effective potential including the  centrifugal term. 
In this framework, the phase $S_{\kappa\ell}(r)$ is approximated as the action (also known as the Hamilton characteristic function~\cite{Goldstein})
\begin{equation}\label{eq:WKBaction}
    S_{\kappa \ell} (r) \simeq  S_{\kappa\ell} (r_0) + \int_{r_0}^r \sqrt{\kappa^2 - 2 V_{\ell} (r')} \; \rmd r' ,
\end{equation}
\end{subequations}
where the value $S_{\kappa\ell} (r_0)$ at the arbitrary radius $r_0$ is enforced to fulfill the asymptotic boundary conditions given in Eq.~\eqref{eqn:Sasym}.

In order to reveal the signatures of the $\ell$-dependent centrifugal term in the two-photon matrix elements, we explicitly express the effective potential felt by the photoelectron with momentum $\ell$ as:
\begin{eqnarray}
\label{eqn:Veff}
    V_{\ell} (r) &=& - \dfrac{Z}{r} + \dfrac{\ell (\ell+1)}{2 r^2} + o(r^{-3}) .
\end{eqnarray}
With the WKB expressions~\eqref{eq:WKB_formalism} and neglecting the $o(r^{-3})$ terms in Eq.~\eqref{eqn:Veff}, we find that the leading term when expanding the WKB action $S_{\kappa\ell}(r)$ in powers of $r^{-1}$ is the first-order term, \ie 
\begin{eqnarray}
\label{eq:DS}
\Delta S_{\kappa \ell}(r) &=& \dfrac{\Re \beta_{\kappa \ell}}{r} + o (r^{-2})
\end{eqnarray}
with the {\em complex valued}, $\ell$-dependent coefficient
\begin{equation}
\label{eq:beta_coeff}
    \beta_{\kappa \ell} = \dfrac{1}{2 \kappa^3}  \left(  \ell (\ell+1) \kappa^2 +  Z^2 + \rmi \, Z \, \kappa \right),
\end{equation}
the imaginary part of which comes from the expansion of the denominator in Eq.~\eqref{eq:WKB_formalism}.
The first term is related to the short-range centrifugal potential $\ell(\ell+1)/(2r^2)$, the last two ones are imprints of the Coulomb tail $-Z/r$.

Eventually, we obtain WKB approximations of the radial continuum  wave functions, up to the first order in $r^{-1}$, behaving  asymptotically as
\begin{subequations}
\label{eqn:Rkl_rhokl_asymptotic}
\begin{eqnarray} \nonumber 
   R_{kL} (r) &=& \dfrac{1}{\sqrt{k}} \rme^{- \Im \beta_{kL}/r} \sin  \big(  S_{kL}^{\infty} (r) + \Re \beta_{kL}/ r \big) + o (r^{-2}), \\ \label{eqn:psiasympt} \\
    \rho_{k_\alpha 1}(r) &=& \dfrac{1}{\sqrt{k_\alpha}} \rme^{- \Im \beta_{k_\alpha 1}/r} \rme^{\rmi \big( S_{k_\alpha 1}^{\infty} (r) + \Re \beta_{k_\alpha 1} / r \big)} + o (r^{-2}) . \nonumber \\ 
\end{eqnarray}
\end{subequations}
Note that, at zeroth order in $r^{-1}$, expressions~\eqref{eq:WKB_formalism} lead to the standard wave functions used in Ref.~\cite{Klunder2011}.

To highlight the importance of the $\Delta S_{\kappa\ell}(r)$ corrections at  short and long distances, we assess our approach with the hydrogen atom, for which exact results are available~\cite{Bethe1957} [in particular, $\eta_{\ell}(k)=\Gamma(\ell+1+\rmi Z/k$)]. 
As an illustration, we show in Fig.~\ref{fig:WKB}a and Fig.~\ref{fig:WKB}b the $L=0$ and $k=0.37$ a.u. radial wave functions. The numerically exact solution of the time-independent Schr\"{o}dinger equation is shown as a blue solid line, the WKB approximation given in Eq.~\eqref{eq:WKB_psi} as a black dashed line and the lowest order equivalent, corresponding to  $\beta_{\kappa \ell}=0$ as a red dotted line. 
We see in frame (b) that both approximated wave functions converge asymptotically to the exact solution. However, our WKB continuum wave function converges much faster to the exact solution due to its $r^{-1}$ corrections, as seen in the shorter range displayed in frame (a). Here, we see that the imaginary part of $\beta_{\kappa \ell}$ modulates the amplitude of the wave function in the short range, which qualitatively reproduce the behavior of the exact wave function~\cite{Dahlstrom2012}. That modulation is formally absent in the zeroth-order wave function.

For a more comprehensive insight, we plotted in Fig.~\ref{fig:WKB}(c,d) the phase difference between each approximate wave function and the exact one (same color code). We see in frame (c) that including the correction allows a much faster convergence of the phase, within few tens of a.u. At $r=40$ a.u, the error is of $-0.015$ rad only when $\beta_{\kappa \ell}$ is properly included, while it is an order of magnitude larger ($0.218$ rad) when enforcing $\beta_{\kappa \ell} =0$. Moreover, frame (d) underlines the improvement of our approach with respect to the zeroth-order approximation even at larger distances, where the latter converges slowly (in $r^{-1}$) while the former accurately matches the exact phase. This is particularly important in the context of the present study because the integrals involved in the matrix element computation are mainly driven by the long-range oscillations of $\rho_{k_\alpha 1}(r) \propto R_{k_\alpha 1}(r)$ and of $R_{k L}(r)$, see below. 

More generally, the phases $S_{\kappa\ell}$ play a central role in interferometric schemes such as the RABBIT. In the next section, we show how the corrections $\Delta S_{\kappa\ell}$ impact the angular dependency of the measurable atomic delays. 

\subsection{Anisotropy of the atomic delay}
We now use our approximate representation of the continuum wave functions to evaluate the two-photon matrix elements involved in the RABBIT process, their phases and the corresponding atomic delay. 
By substituting Eqs.~\eqref{eq:WKB_formalism} into~\eqref{eq:Talphaell}, and neglecting  the phase term proportional to $S_{kL} {+} S_{k_{\alpha}1}$ following Ref.~\cite{Klunder2011} (rotating-wave approximation), we obtain 
\begin{equation}\label{eq:bite}
     T^{\alpha}_{L} (k) \simeq \int_{0}^{\infty}\rmd r  \dfrac{r \exp \big( \rmi \big[ S_{kL} (r)-S_{k_\alpha 1} (r) \big] \big) }{[ k^2 {-} 2 V_{L} (r) ]^{1/4} [ k_\alpha^2 {-} 2 V_{1} (r) ]^{1/4}}  .
\end{equation}
By approximating $S_{\kappa\ell}(r)$ by $S_{\kappa\ell}^\infty(r){+}\Re \beta_{\kappa \ell}/r$ and $[\kappa^2{-}2V_\ell(r)]^{-1/4}$ by $\exp(-\Im \beta_{\kappa \ell}/r)/\sqrt{\kappa}$ consistently with Eqs.~\eqref{eqn:Rkl_rhokl_asymptotic}, we obtain an analytic expression for the radial two-photon transition matrix elements
\begin{subequations}
\label{eq:TM_2photon}
\begin{eqnarray}
    && T_{L}^{\alpha} (k) \simeq \dfrac{-\rmi^{L}}{\sqrt{k \, k_{\alpha}}} \rme^{\rmi [ \eta_1 (k_\alpha)- \eta_L (k) ]}  \nonumber \\ 
    && \qquad \times \frac{(2k_{\alpha})^{\rmi Z/k_{\alpha}}}{(2k)^{\rmi Z/k}}  \; \mathbb{M} \Big( \underbrace{Z/k_{\alpha} - Z/k}_{\equiv s}, \underbrace{k_{\alpha} - k}_{\equiv \Delta k} , \underbrace{\beta_{k_{\alpha}1}  - \beta_{kL}^{\ast}}_{\equiv \Delta \beta_L} \Big) , \nonumber \\ \label{eq:T_2photon}
\end{eqnarray}
with
\begin{eqnarray}
    \mathbb{M} (s,\Delta k , \Delta \beta_L ) &\equiv& \dfrac{1}{2} \int_0^{\infty}r^{1+\rmi s}\exp\left(\rmi \Delta k \, r + \rmi \frac{\Delta \beta_L}{r}\right) \; \rmd r \nonumber\\
    &=& \left( \dfrac{\Delta\beta_L}{\Delta k} \right)^{1+\frac{\rmi s}{2}}  {\rm K}_{2{+}\rmi s} \Big( - 2 \rmi \sqrt{\Delta \beta_L \, \Delta k} \Big) ,\nonumber \\ \label{eq:M_2photon}
\end{eqnarray}
\end{subequations}
where ${\rm K}_{\nu} (z)$ is the modified Bessel function of the second kind. Note that in Ref.~\cite{Dahlstrom2012} the correction in $r^{-1}$ has been considered only in the modulus  of the continuum wavefunction  (\ie corresponding to taking the real part of $\beta_{\kappa\ell}$ to zero).

The atomic phase $\tau_{A} (\mathbf{k})$ can then be evaluated by inserting the analytical expression~\eqref{eq:T_2photon} in Eq.~\eqref{eq:Mtot}. Expression~\eqref{eq:T_2photon} shows explicitly that  $\eta_1 (k_e) {-} \eta_1 (k_a)$ emerges and $\eta_L(k)$ disappears when expressing $\Delta\phi_A(\mathbf{k})$ out of $T_{L}^{\alpha} (k)$, as in~\cite{Klunder2011}.  
Here, $\tau_A(\mathbf{k})$ and therefore $\taucc (\mathbf{k})$ are anisotropic due to the explicit $L$-dependence of  $\arg T_L^{\alpha}$. 
We now turn to the comparison of the results with {\em ab initio} numerical calculations.

\begin{figure}
    \centering
    \includegraphics[width=.5\textwidth]{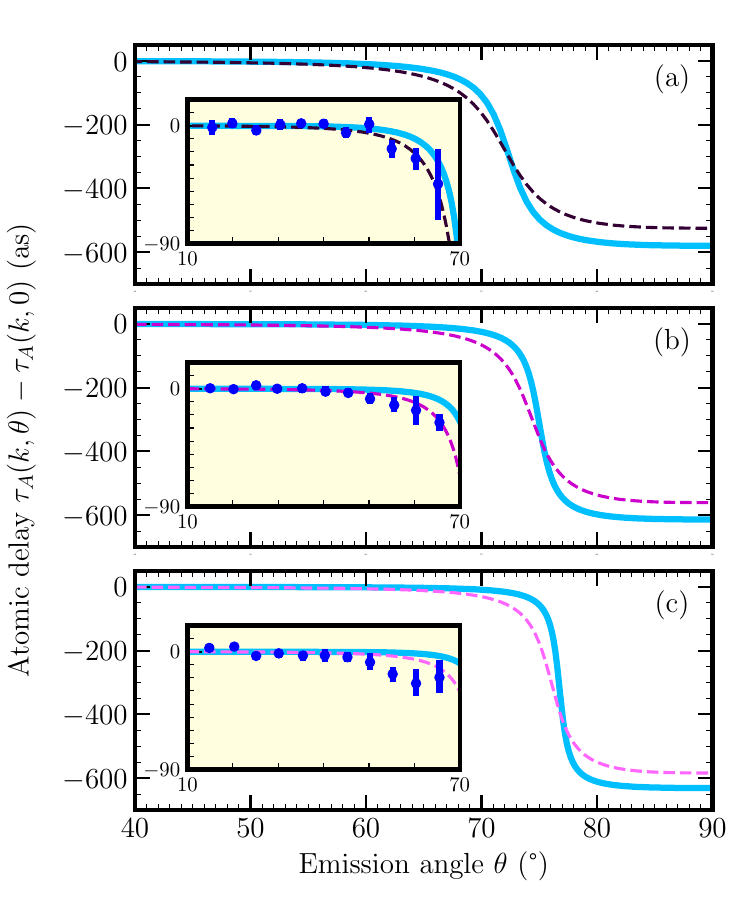}
     \caption{Angular variations of the atomic delay, $\tau_A(\theta)-\tau_A(0^{\circ})$, in photoemission from a neutral atom. Data obtained from the numerically exact simulations on the H atom~\cite{Cionga1993} (solid lines), using the analytical WKB derivation with $Z=1$ from Eqs.~\eqref{eq:TM_2photon} (dashed lines) and experimentally on He  reproduced from~\cite{Heuser2015} (blue dots with error bars). Each panel is associated with a given asymptotic final momentum (see Fig.~\ref{fig:rabbitscheme}): $k=0.61$ a.u (a), $k=0.77$ a.u (b) and $k=0.91$ a.u (c).}
     \label{fig:RABBIT_phase}
 \end{figure}

Figure~\ref{fig:RABBIT_phase} shows the angular variations of the atomic delay, $\Delta\tau_A(k,\theta)=\tau_A(k,\theta)-\tau_A(k,0)$, as a function of the emission angle $\theta$ with respect to the polarization axis,  for few  values of the asymptotic final momentum $k$.  The WKB results obtained from Eq.~\eqref{eq:T_2photon} are shown with dashed lines and the ones obtained from numerically exact calculations in solid lines. 
All the curves follow a similar trend. First, $\Delta\tau_A$ remains constant, close to zero, for $0\degree<\theta\lesssim 60\degree$ at all $k$. Then a sudden jump takes place, spanning several 100 attoseconds over few 10\degree, before reaching a stable value for larger $\theta$ values. Besides, we make  the two following observations when $k$ increases: (i) the critical angle $\theta_c$ around which the jump occurs increases; (ii) the jump gets sharper and its magnitude converges towards $\sim\pi/(2\wL)$.
The latter is the signature of a change of sign in the dominant  contribution $M_a (\mathbf{k})$ to the overall amplitude~\cite{Busto2019}. In the context of RABBIT measurements of one-photon ionization delays, this remarkable \wL-dependent feature underlines the  probe-origin of the measured $\tau_A$ anisotropy~\cite{Heuser2015}. 

These $k$- and $\theta$-dependent trends, observed in the exact simulations (thick blue lines), are qualitatively reproduced by our analytical WKB results (dashed lines). Quantitatively, we note that $\theta_c$  are overestimated in the analytical predictions by about $5^{\circ}$ compared to the  exact solutions and the $\Delta\tau_A$ jump is less pronounced than in the exact simulations by about $50$~as.
These small differences can be related to the estimation of the WKB wave functions near the origin [see Fig.~\ref{fig:WKB}(a)].
Nevertheless, these results show that the approximate WKB wave functions expanded to the first order in $r^{-1}$ are sufficient to recover the  trend of the experimental results reported in~\cite{Heuser2015}, displayed as symbols with errorbars in the insets of Fig.~\ref{fig:RABBIT_phase}. 

Therefore, our results show that the angular dependency of $\taucc$  mainly comes from a $r^{-1}$ contribution to the photoelectron phase. This term plays a significant role because it persists at large distances. It is itself the signature of the short-range influence of the centrifugal barrier, which indicates that the IR pulse, in the RABBIT setup, probes the photoelectron already at short distances.

\subsection{Fano's propensity rules \label{sec:soft-photon}}

We now revisit the anisotropy of RABBIT measurements in terms of Fano's propensity rules~\cite{Fano1985,Busto2019}, in the light of our WKB-based approach. 

In order to obtain an expression of \taucc bearing relevant physical insight, we proceed with our analytical developments in the so-called soft-photon limit~\cite{Maquet2007}, which is achieved either for large $k$ or small $\wL$. The soft-photon approximation has often proved efficient in this context, see \eg~\cite{Dahlstrom2013, Bray2018}.
This corresponds to considering $\Delta k/k$  small  in the computation of the matrix elements expressed according to Eq.~\eqref{eq:T_2photon} and~\eqref{eq:M_2photon}, leading to 
\begin{equation} \label{eq:EqMsoftphoton}
    \mathbb{M} (s,\Delta k , \Delta \beta_L ) \approx \dfrac{\Gamma (2+\rmi s)}{2 (- \rmi \Delta k)^{2+\rmi s}} \exp \left( \dfrac{\Delta k \; \Delta \beta_L}{1+\rmi s} \right),
\end{equation}
where we have used standard limits of Bessel functions~\cite{Abramowitz1974}.
One should note that, the isotropic results of~\cite{Klunder2011, Dahlstrom2013} can  be retrieved by  setting $\Delta \beta_L = 0$ {\em a posteriori} in this last expression of the matrix element.

Next, by substituting Eq.~\eqref{eq:EqMsoftphoton} into Eq.~\eqref{eq:T_2photon} and  expanding in terms of $\wL / k^2$ up to the first order, we obtain the following expression for the radial transition matrix element in the soft-photon regime
\begin{eqnarray} 
  T^{\alpha}_{L}(k) &\approx& \dfrac{\rmi^L k}{2 \wL^2} \rme^{\rmi [ \eta_1 (k_\alpha)- \eta_L (k) ]} \exp \left( \epsilon_{\alpha} \dfrac{\wL}{2 k^2} [2 - L (L+1)] \right)  \nonumber \\
  &\times&  \exp\left(\rmi \epsilon_{\alpha}  \frac{ \wL Z}{k^3} \left[ 1 +  \gamma + \ln\left( \frac{\wL}{2k^2}\right)\right]\right) , \label{eq:Tsoftphoton}
\end{eqnarray}
where $\epsilon_a={-}1$ and $\epsilon_e = {+}1$,  $\gamma \approx 0.577$ is the Euler-Mascheroni constant. Remarkably, at this level of approximation $\arg T^{\alpha}_{L}(k)$ does not depend on the final electron momentum $L$ other than through the final scattering phase $\eta_L (k)$ (and $\rmi^L$), in contrast to its modulus 
\begin{equation}
    | T^{\alpha}_{L}(k)| = \dfrac{k}{2\omega_0^2} \exp \left(  \dfrac{\epsilon_{\alpha} \omega_0}{2 k^2} [2 - L(L+1)] \right).
\end{equation}
This expression allows retrieving, in the soft-photon limit, the Fano propensity rules~\cite{Busto2019} of two-photon processes.
Indeed, it tells us that $|T_2^{a}| \geq |T_0^{e}| \geq |T_0^{a}| \geq |T_2^{e}|$, \ie the most probable ionization path in the RABBIT scheme is the one involving the IR absorption ($\alpha=a$) with $L=2$ while the least probable path is the one involving the IR emission ($\alpha=e$) for $L=2$. These propensity rules appear to result from the $L$-dependency of the real part of the lowest order in $\wL/k^2$ of 
\begin{equation}
\label{eq:beta0_coeff}
 \Delta \beta_L = \dfrac{2 - L(L+1)}{2 k} + \rmi \dfrac{Z}{k^2} + o \left( \frac{\wL}{k^2} \right) ,
\end{equation}
and therefore from the $r^{-1}$ term [Eq.~\eqref{eq:DS}] in the continuum wave function phase [Eq.~\eqref{eqn:S}]. 

In terms of dynamics, the atomic delay $\tau_A$ in the soft-photon regime is obtained by substituting Eq.~\eqref{eq:Tsoftphoton} into Eqs.~\eqref{eq:Mtot} and~\eqref{eq:tau_RABBIT}, leading to the expression
\begin{subequations}
\label{eq:soft-photon}
\begin{equation}
    \lim_{\wL \to 0} \taucc(\mathbf{k}) = \frac{Z}{k^3}\left[ 1+\gamma+\ln\left(\frac{\wL}{2k^2}\right)\right] - \dfrac{1}{2\wL} \arg \big( f(\mathbf{k})\big) , 
\end{equation}
with the real-valued, orientation-dependent, function
\begin{equation}\label{eq:ftheta}
    f(\mathbf{k}) =1 -  \dfrac{2 \; C_{00}  Y_{00} (\hat{k}) \; C_{20} Y_{20} (\hat{k})}{[C_{00}Y_{00}(\hat{k}) ]^{2} + [C_{20}Y_{20}(\hat{k}) ]^{2}} \cosh \left( \dfrac{3 \wL}{k^2} \right)  ,
\end{equation}
\end{subequations}
for all $Z$.
The angular jump observed in each frame of Fig.~\ref{fig:RABBIT_phase} is thus the signatures of a change of sign in $f$. This can occur only due to the $\cosh$ factor, the presence of which being a direct manifestation of the Fano propensity rules. Indeed, this factor reduces to 1 if one sets $|T_2^{a}| = |T_0^{a}|$ and $|T_0^{e}| = |T_2^{e}|$, which is underlied by the original derivations of Refs.~\cite{Klunder2011,Dahlstrom2013}, see Appendix~\ref{sec:Klunder}.

\begin{figure}
    \centering
    \includegraphics[width=.4\textwidth]{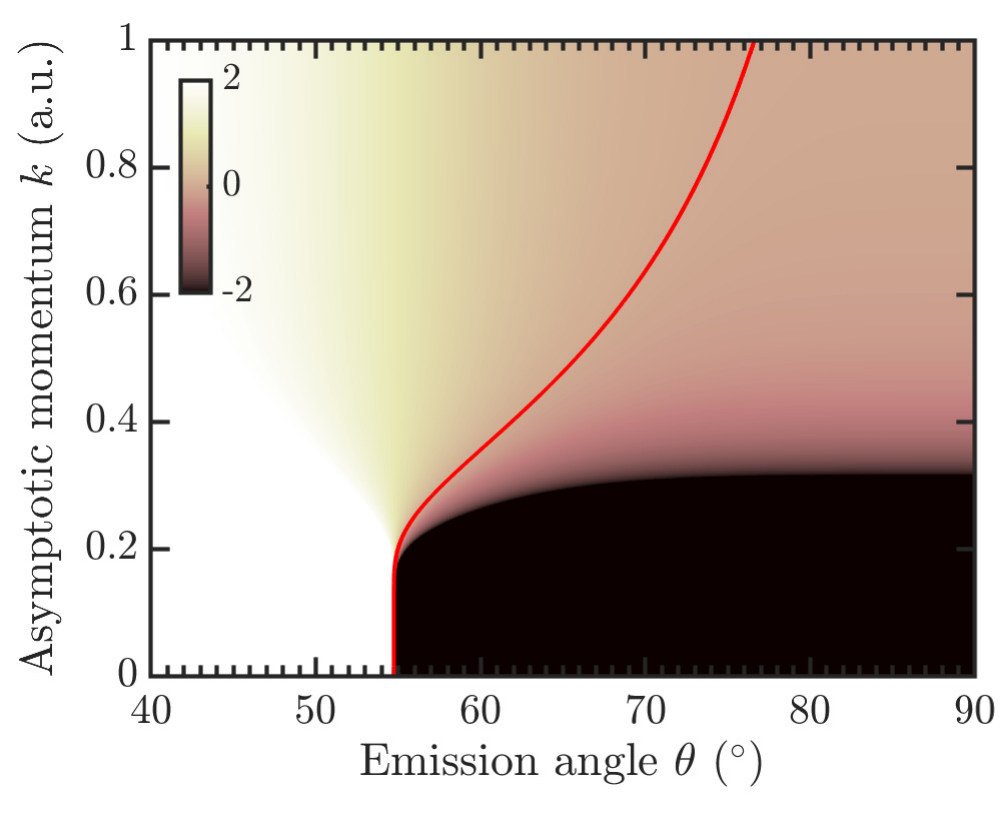}
    \caption{Value of $f(\mathbf{k})$ given by Eq.~\eqref{eq:ftheta} as a function of the asymptotic momentum $k$ and the emission angle $\theta$. The red line indicates $f(\mathbf{k}) = 0$.}
    \label{fig:ftheta}
\end{figure}

The angle at which the $\pi$-jump occurs increases for increasing $\wL/k^2$, as indicated by the red line in the $(k,\theta)$ map of Fig.~\ref{fig:ftheta}. This is consistent with the general trend illustrated in Fig.~\ref{fig:RABBIT_phase} and commented above.
Note that higher orders in the soft-photon expansions would be required to investigate the \wL-dependency of  $\taucc(k,180\degree)-\taucc(k,0\degree)$ revealed numerically on 1D {\em asymmetric} model systems in~\cite{Bekane2024}. 

\section{Links with available derivations and interpretations}\label{sec:links}
Here, we highlight additional links between our semi-analytical approach and existing derivations and interpretations, and exploit the advantages of the WKB approach to gain further physical insights. We resort to the soft-photon approximation invoked in Section~\ref{sec:soft-photon} when it is helpful.

\subsection{Orientation-averaged RABBIT}
\begin{figure}
    \centering
      \includegraphics[width=.45\textwidth]{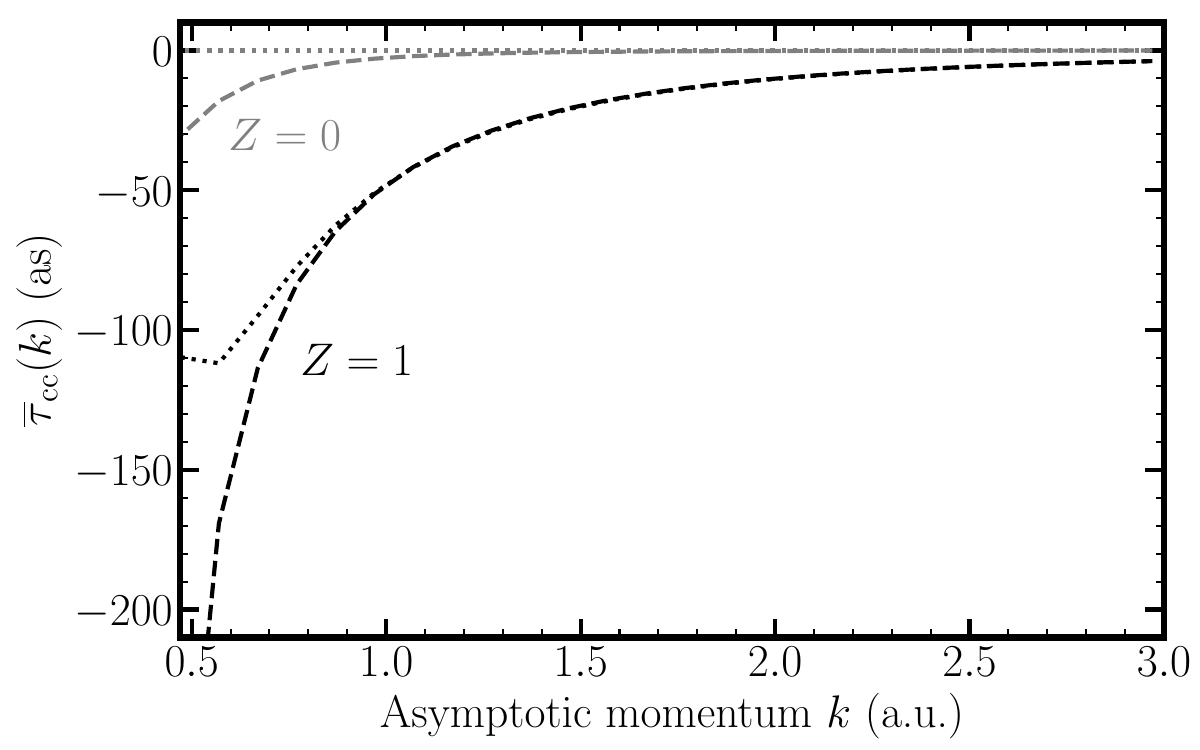}
     \caption{Continuum-continuum time delays $\overline{\tau}_{\rm cc} (k)$ from the integrated RABBIT scheme as a function of the asymptotic momentum $k$. The grey and black lines are for $Z=0$ and $Z=1$, respectively. The dashed lines show the data obtained with the radial two-photon transition matrix element~\eqref{eq:T_2photon}. The dotted lines correspond to the soft-photon approximation given by Eq.~\eqref{eq:soft-photon-integrated}.}
     \label{fig:tauAphotodetachement}
 \end{figure}

In the original RABBIT scheme~\cite{Paul2001}, the atomic phase $\overline{\Delta \phi}_A (k)$ is extracted from the orientation-averaged photoelectron spectra
\begin{equation}
    \overline{\cal I}(k;\tauxuvir) = \int \mathcal{I}(\mathbf{k};\tauxuvir) \;  \rmd \hat{k} .
\end{equation}
The atomic delay obtained in this context is then given by
\begin{equation}\label{eqn:phiAintegre}
    \overline{\tau}_A (k) = -\frac{1}{2\wL} \arg \left( \sum_{L=0,2} |C_{L0}|^2 T_L^{a} (k) T_L^{e} (k)^{\ast}  \right) .
\end{equation}
In the same way as for the angularly-resolved case,  the angle-integrated continuum-continuum correction  is defined as 
$\overline{\tau}_{\rm cc} (k) = \overline{\tau}_A (k) - \overline{\tau}_W (k)$.  
Note that, if the probed one-photon transition ends-up in a single angular momentum channel as considered throughout the paper, $\overline{\tau}_W (k)=\tau_W (\mathbf{k})$ in all momentum directions. Otherwise, if it ends-up in a combination of angular momenta, \ie starting from a $\ell{\neq}0$ initial state, then the Wigner delay $\overline{\tau}_W (k)$ associated with an orientation-averaged  measurement is ill-defined.
In the soft-photon limit, we find that
\begin{equation}
\label{eq:soft-photon-integrated}
    \lim_{\wL \to 0} \overline{\tau}_{\rm cc} (k) =\frac{Z}{k^3}\left[ 1+\gamma+\ln\left(\frac{\wL}{2k^2}\right)\right] .
\end{equation}
This result corresponds to the isotropic contribution of the angularly resolved case given by Eq.~\eqref{eq:soft-photon}. We have verified that Eq.~\eqref{eq:soft-photon-integrated} can be derived by taking the soft-photon limit of Eq.~(100) in Ref.~\cite{Dahlstrom2012}. Figure~\ref{fig:tauAphotodetachement} compares $\overline{\tau}_{\rm cc}$ obtained from the radial two-photon transition matrix element [Eq.~\eqref{eq:T_2photon}] (dashed lines) and within the soft-photon approximation [Eq.~\eqref{eq:soft-photon-integrated}] (dotted lines), for $Z=1$ (black lines) as well as for $Z=0$ (grey lines).  We observe that for all $k\gtrsim 1$,  the soft-photon approximation provides a very good estimate of the actual $\overline{\tau}_{\rm cc}$.  Note that, when neglecting the long range term (in $r^{-1}$) in the wave function, the resulting expression significantly differs from Eq.~\eqref{eq:soft-photon-integrated}, see  Appendix~\ref{sec:Klunder}.

The expression~\eqref{eq:soft-photon-integrated} is reminiscent of the one obtained from empirical classical arguments in Ref.~\cite{Pazourek2015}, \ie
\begin{equation}
\label{eq:tauccPazourek}
    \overline{\tau}_{\rm cc} (k) \approx \frac{Z}{k^3}\left[ 2 + \ln \left(\frac{\wL}{\pi k^2}\right) \right].
\end{equation} 
It was shown to reproduce well {\em ab initio} calculations in the soft-photon regime~\cite{Bray2018,Bekane2024}.
The absolute difference between the expressions from Eqs.~\eqref{eq:soft-photon} and~\eqref{eq:tauccPazourek} is  $(\gamma-1+\ln(\pi/2))Z/k^3 \simeq 0.03 {\times}Z/k^3$ which  indeed vanishes far from threshold.

\subsection{Atomic delays in photodetachment processes \label{sec:photodetachment}}
We now consider photodetachment processes from closed-shell negative ions, as investigated numerically in Ref.~\cite{Lindroth2017} from the delay perspective. In this case, the effective potential is given by Eq.~\eqref{eqn:Veff} with $Z=0$. It is short-range and asymptotically dominated by the centrifugal barrier. 
As a consequence, the radial dependency of the asymptotic phase $S^{\infty}_{\kappa \ell} (r)$ of the continuum wave function in Eq.~\eqref{eqn:S} boils down to $\kappa r$, which corresponds to a free particle.
Regarding the coefficient of the correcting term $\Delta S_{\kappa \ell}(r)$ at order $r^{-1}$  [Eq.~\eqref{eq:DS}], it becomes  $\beta_{\kappa \ell} = \ell (\ell {+} 1) / 2 \kappa$. Therefore, even in the presence of short-range interactions only, the phase maintains a long-range behavior.

In order to analyse the photodetachment dynamics, we start by comparing the continuum-continuum time delays from the integrated RABBIT scheme shown in Fig.~\ref{fig:tauAphotodetachement} and discussed in the previous Section. 
For increasing $k$, $\overline{\tau}_{\rm cc} (k)$ goes faster to zero  for $Z=0$ than for $Z=1$, and the photodetachment time-delay practically vanishes beyond $k\approx 1$, in agreement with Ref.~\cite{Lindroth2017}. This is consistent with the hypothesis according to which, in a RABBIT experiment, the interaction with the IR photon takes place outside the effective range of the atomic potential, including the centrifugal term, leading to $\overline{\tau}_A (k) \approx \overline{\tau}_W (k)$.  

\begin{figure}
    \centering
    \includegraphics[width=0.5\textwidth]{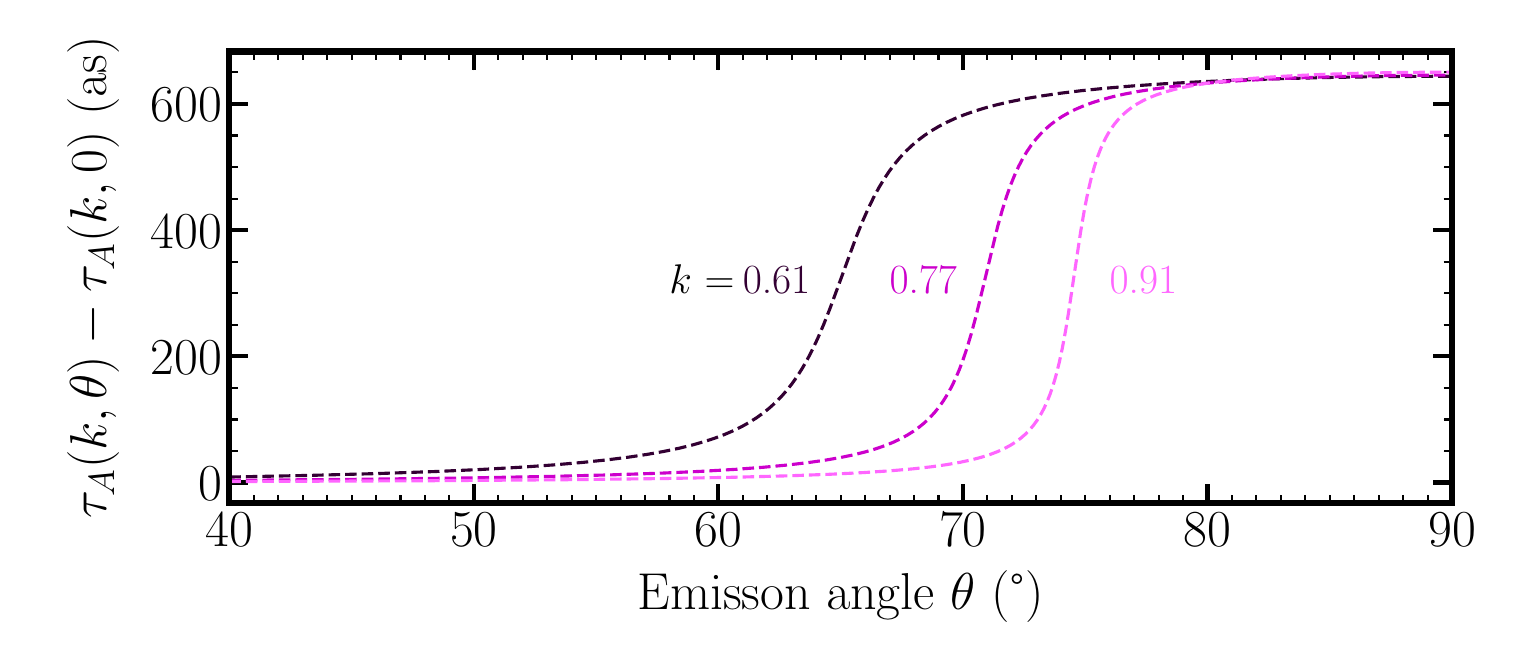}
     \caption{Angular variations of the atomic delay, $\tau_A(k,\theta)-\tau_A(k,0\degree)=\taucc(k,\theta)-\taucc(k,0\degree)$, in a photodetachement process. Data obtained using Eqs.~\eqref{eq:TM_2photon} with $Z=0$. The dark violet, medium purple and light pink lines correspond to values of the final asymptotic momentum $k=0.61$ a.u, $0.77$ a.u. and $0.91$ a.u, respectively. }
    \label{fig:photodetachement}
\end{figure}
Meanwhile, when looking at the angularly resolved data, see Fig.~\ref{fig:photodetachement}, we observe an angular dependency which is as pronounced as in the photoionization case ($Z=1$), see Fig.~\ref{fig:RABBIT_phase}. Indeed, as explained in Section~\ref{sec:soft-photon}, the main features of this angular dependency are related to the universal centrifugal barrier. Therefore, the IR transitions occur under the influence of the latter also in the photodetachment case, even though it does not manifest in the angularly integrated data.

\subsection{Classical perspective on the Fano propensity rules}

In this final section, we exploit the WKB formalism to pinpoint the dominant paths in the two-photon process  and highlight the classical mechanisms behind Fano's propensity rules discussed in Section~\ref{sec:soft-photon}.

\begin{figure*} 
    \centering
    \includegraphics[width=.8\textwidth]{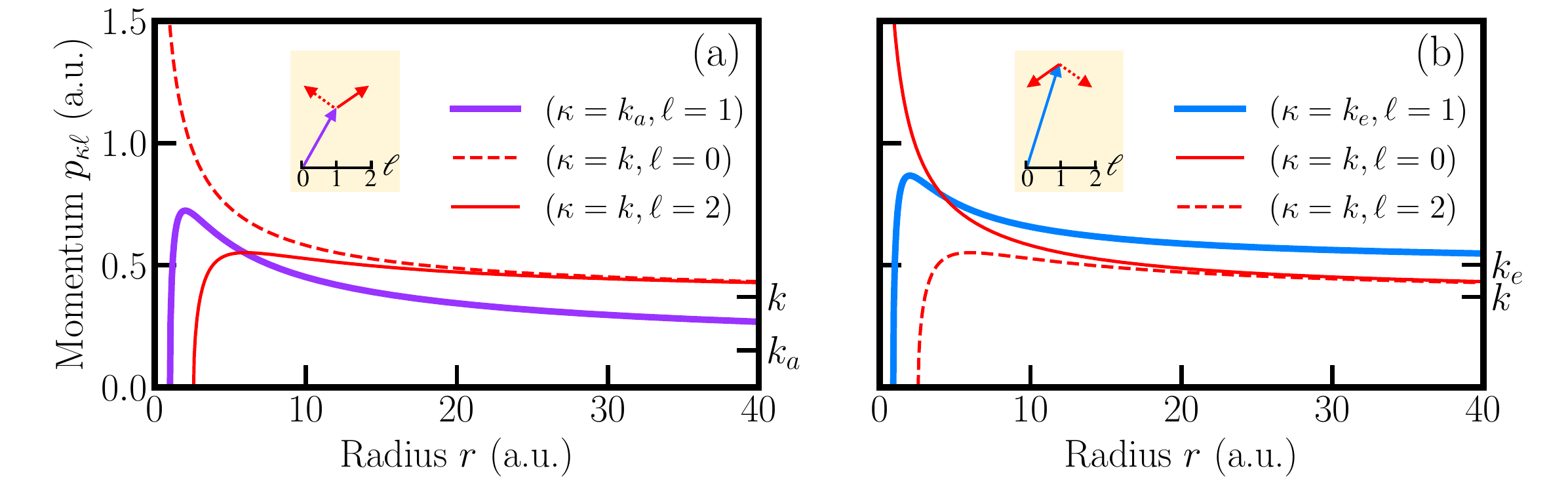}
     \caption{Local momentum $p_{\kappa\ell}$ Eq.~\eqref{eqn:localmomentum} as a function of the radius in intermediate states ($\ell=1$) and final states ($\ell=0,2$) involved in the two arms of the interferometric 800-nm RABBIT scheme (left and right panels respectively). The considered asymptotic final momentum is $k=0.37$, such that the asymptotic intermediate  momenta are $k_a=0.15$~a.u. et $k_e=0.50$~a.u. }
     \label{fig:spacephase}
 \end{figure*}

We start with the WKB expression of $T^{\alpha}_{L}(k)$ given by Eq.~\eqref{eq:bite} and note that 
 it is reminiscent of the single-photon form discussed by Fano in Ref.~\cite{Fano1985}.
Its integrand contains an oscillating complex term, $\exp (\rmi (S_{kL} {-} S_{k_{\alpha}1}))$.
Following Fano's arguments, and inspired by the saddle-point approximation, the modulus of this integral is larger if there exists a real-valued position for which the derivative of the phase vanishes. 
More explicitly, this corresponds to the conditions 
\begin{equation} \label{eqn:saddle}
    p_{k L} = p_{k_{\alpha}1} ,
\end{equation}
where 
\begin{equation}\label{eqn:localmomentum}
    p_{\kappa \ell} (r)\equiv \frac{\partial S_{\kappa\ell} (r)}{\partial r} =\sqrt{\kappa^2 - 2V_\ell(r) }
\end{equation}
is the local momentum obtained from the Hamilton characteristic function $S_{\kappa \ell}$ given in Eq.~\eqref{eq:WKBaction}~\cite{Goldstein}. 

We now turn to the identification of the paths, among those involved in a RABBIT scheme, fulfilling  Eq.~\eqref{eqn:saddle}  or not. As an illustration, we consider the $Z=1, k=0.37$ case. Figure~\ref{fig:spacephase} shows trajectories $(r,p_{\kappa \ell})$ in a phase-space-like representation, for the different $\kappa$ and $\ell$ associated with the intermediate and final states. The `absorption' and `emission' paths are shown in Fig.~\ref{fig:spacephase}a and  Fig.~\ref{fig:spacephase}b, respectively.
In each frame, the  $(r,p_{k_{\alpha}1})$ and $(r,p_{kL})$ trajectories are represented by thick and thin lines, respectively. The existence of a real-valued saddle point fulfilling~\eqref{eqn:saddle} is therefore revealed by an intersection between thick and thin lines. Note that the trajectories can possibly intersect because they are associated with different effective potentials $V_{\ell}$.

The trajectories possessing such a point are highlighted in solid lines, while the others are shown with dashed lines. In Fig.~\ref{fig:spacephase}a ($\alpha=a$), only the trajectory $(r,p_{k 2})$ intersects with $(r,p_{k_a 1})$, and therefore the dominant quantum path is the $L =2$ channel. 
In Fig.~\ref{fig:spacephase}b ($\alpha=e$), only the trajectory $(r,p_{k 0})$ intersects with $(r,p_{k_e 1})$, and therefore the dominant quantum path is the $L =0$ channel. 
Hence, this perspective expands  the classical arguments  implied by Fano's discussion on the propensity rules in Ref.~\cite{Fano1985} to the two-photon case~\cite{Busto2019}. 

\section{Conclusions}\label{sec:conclusion}

To summarize, we have used the WKB formalism to obtain approximate atomic continuum wave functions that provide a qualitative account of Fano's propensity rules in two-photon ionization~\cite{Fano1985}. The latter are responsible for the probe-induced asymmetry in interferometric RABBIT measurements of the Wigner delay characterizing the one-photon ionization dynamics~\cite{Heuser2015,Bray2018,Busto2019}. 

Our derivations provide analytical expressions of the radial two-photon transition matrix elements [Eq.~\eqref{eq:Tsoftphoton}] and of the so-called atomic delay $\tau_A$ [Eqs.~\eqref{eq:soft-photon}], in the soft-photon regime. It notably accounts for the angular jump of nearly $\pi$ rad in the argument of the RABBIT transition matrix element, due to Fano's propensity rules. This jump leads to a pronounced anisotropy of the atomic time delay measurements (\ie an angular jump of approximately $\pi/(2\wL)$ in terms of delay). We have shown that this probe-induced asymmetry of $\tau_A$ can be traced back to a universal long range behavior, in $r^{-1}$, of the continuum wave function, which is inherited from the long range Coulomb tail of the ionic potential as well as from the $\ell$-dependent short-range centrifugal potential. 

We then further investigated how our approach relates to other previously published results. When considering orientation-averaged RABBIT measurements, we have obtained an analytical expression of the atomic time delay $\overline{\tau}_A$ [Eq.~\eqref{eq:soft-photon-integrated}] that is remarkably close to the one obtained semi-empirically in Ref.~\cite{Pazourek2015} and that has proved efficient in comparisons with numerical simulations~\cite{Bray2018, Berkane2024}. We have also investigated the anisotropy of the atomic delay in photodetachement processes, previously studied numerically in Ref.~\cite{Lindroth2017} from the orientation-averaged perspective. Our results show that the anisotropy is as pronounced as in the photoionization case, since it appears as an imprint of the universal centrifugal potential on the scattering phase.

Eventually, we exploited the classical insight offered by the WKB approximation to highlight an intuitive interpretation of the Fano's propensity rules in terms of momentum conservation~\cite{Fano1985}, by evaluating the transition matrix elements with the saddle-point approximation. The simplicity of the analytical formulas obtained here offers a promising framework for the investigation of photoemission time delays in atomic systems from initial states carrying a non-zero orbital quantum number~\cite{Cirelli2018,Hockett2017}, and more generally in systems with additional degrees of freedom~\cite{Worner2016,Vos2018,Khademhosseini2023,Boyer2024} and beyond the RABBIT regime~\cite{Trabert2023}.

\section*{Acknowledgements}

This research received financial support from the French
866 National Research Agency through Grants No. ANR-20-CE30-0007-868-DECAP and No. ANR-21-CE30-0036-03-ATTOCOM.

\appendix

\section{Zeroth-order derivations and predictions \label{sec:Klunder}}
In this appendix, we recall the theory of Ref.~\cite{Klunder2011, Dahlstrom2012, Dahlstrom2013} to approximate the radial two-photon transition matrix elements in Eq.~\eqref{eq:Talphaell}. We then show the predictions it leads to in terms of photoemission time delays, regardless of the WKB approximations used in the present paper.

Using the paper's notations, the continuum wave functions in Eq.~\eqref{eq:Talphaell} are approximated by asymptotic Coulombic wave functions 
\begin{subequations}
\label{eqn:Rkl_rhokl_asymptotic_Klunder}
\begin{eqnarray} 
   R_{kL} (r) &=& \dfrac{1}{\sqrt{k}} \sin  \big(  S_{kL}^{\infty} (r) \big) + o(r^{-1}), \label{eqn:psiasympt_Klunder} \\
    \rho_{k_\alpha 1}(r) &=& \dfrac{1}{\sqrt{k_\alpha}} \rme^{\rmi  S_{k_\alpha 1}^{\infty} (r) }  + o (r^{-1}) ,  
\end{eqnarray}
\end{subequations}
where the expression of $S_{\kappa \ell}^{\infty} (r)$ are defined in Eq.~\eqref{eqn:S}.
Substituting Eqs.~\eqref{eqn:Rkl_rhokl_asymptotic_Klunder} in Eq.~\eqref{eq:Talphaell} leads to radial two-photon transition matrix elements of the form
\begin{equation}\label{eq:T_Klunder}
    T_L^{\alpha} (k) \simeq \dfrac{- \rmi^{L}}{\sqrt{k \, k_{\alpha}}}  \rme^{\rmi [ \eta_1 (k_\alpha)- \eta_L (k) ]} \frac{(2k_{\alpha})^{\rmi Z/k_{\alpha}}}{(2k)^{\rmi Z/k}} \dfrac{\Gamma (2+\rmi s)}{2(-\rmi \Delta k)^{2+\rmi s}} .
\end{equation}
As a consequence, in this case, all radial two-photon matrix elements are related through
\begin{equation}\label{eq:T_relation_Klunder}
    \rmi^{L} T_L^{\alpha} (k) \, \rme^{\rmi \eta_L (k)} = T_0^{\alpha} (k) \, \rme^{\rmi \eta_0 (k)} ,
\end{equation}
for all $L$. 
The relation~\eqref{eq:T_relation_Klunder} has drastic consequences in the angularly resolved atomic time delay and the one obtained from the integrated RABBIT measurements that are intrinsic to the approximation~\eqref{eqn:Rkl_rhokl_asymptotic_Klunder} that we detail below.
Substituting Eq.~\eqref{eq:T_relation_Klunder} in Eq.~\eqref{eq:Mtot}, we obtain 
\begin{equation}
    M_{\alpha} (\mathbf{k}) = (8 \pi)^{5/2}  \dfrac{\rmi \rme^{\rmi \eta_0 (k)} T_{0}^{\alpha}(k)}{6 \sqrt{k \, k_{\alpha}}} \sum_{L = 0,2} \rmi^L C_{L 0} Y_{L 0} (\hat{k}) .
\end{equation}
The atomic phase from Eq.~\eqref{eqn:atomic_phase} simplifies to
\begin{equation}
    \Delta \phi_A (\mathbf{k}) = \arg \big( T_{0}^{a} (k) \big) - \arg \big( T_{0}^{e}(k) \big) .
\end{equation}
Therefore, it is independent of the asymptotic momentum angle $\hat{k}$, \ie $\tau_W (\mathbf{k})$ and $\taucc (\mathbf{k})$ are isotropic.

We now turn to the atomic phase obtained from the angularly integrated RABBIT scheme whose expression is given in Eq.~\eqref{eqn:phiAintegre}. 
Here again, we use the relation~\eqref{eq:T_relation_Klunder} and we obtain
\begin{equation}
    \overline{\Delta \phi}_A (k) = \arg \big( T_{0}^{a} (k) \big) - \arg \big( T_{0}^{e}(k) \big) .
\end{equation}
Hence, $\Delta \phi_A (\mathbf{k}) = \overline{\Delta \phi}_A (k)$ and $\taucc(\mathbf{k}) = \overline{\tau}_{\rm cc} (k)$.
In the soft-photon limit discussed in Sec.~\ref{sec:soft-photon}, we find in this case 
\begin{equation}
    \lim_{\wL \to 0} \overline{\tau}_{\rm cc} (k) = \dfrac{Z}{k^3} \left[ \gamma + \ln \left( \dfrac{\wL}{2 k^2}\right) \right] ,
\end{equation}
which differs significantly from the empirical formula given by Eq.~\eqref{eq:tauccPazourek} and the one derived in this article~\eqref{eq:soft-photon-integrated}.
Moreover, for the photodetachment described in Sec.~\ref{sec:photodetachment} (\ie for $Z =0$), all terms in the right-hand side of Eq.~\eqref{eq:T_Klunder} become real except the ones coming from the scattering phase. Hence, 
\begin{equation}
    \overline{\Delta \phi}_A (k) = \eta_1 (k_a) - \eta_1 (k_e) , 
\end{equation}
that is the Wigner phase delay and $\taucc (k) =0$ for all asymptotic momenta. In contrast, it is non-vanishing in the numerical simulations~\cite{Lindroth2017} for intermediate $k$.

To conclude, features observed in the numerical simulations~\cite{Bray2018} and the experiments~\cite{Heuser2015} are not present in the approximation used in~\cite{Klunder2011, Dahlstrom2012} and summarized in this appendix, namely: The anisotropy in the continuum-continuum time delay $\taucc (\mathbf{k})$ for all $Z$, the soft-photon regime is not well reproduced, and the non-vanishing continuum-continuum time delay for $Z=0$. These features result from the long-range behavior in $r^{-1}$ of the continuum wave functions, as demonstrated in the main text of this article, and therefore cannot be captured by the approximation~\eqref{eqn:Rkl_rhokl_asymptotic_Klunder}.


%

\end{document}